%% file: Paper.tex
\documentclass[10pt,conference,compsoc]{IEEEtran}

\usepackage[utf8]{inputenc}
\usepackage{booktabs}
\usepackage{amsmath}
\usepackage{amssymb}
\usepackage{xcolor}
\usepackage{url}
\usepackage{graphicx}
\usepackage{orcidlink}
\usepackage{hyperref}
\hypersetup{hidelinks}
\usepackage{balance}

\newcommand{\code}[1]{\texttt{#1}}
\input{tables/tab_numbers.tex}
\input{tables/tab_audio_numbers.tex}
\input{tables/tab_track_numbers.tex}

\begin{document}

\title{Open-Source Intelligence and Music Information Retrieval for Geographic
Attribution of Musical Affect and the Ecological Limits of Population Inference}

\author{
\IEEEauthorblockN{Mohammadreza Rashidi~\orcidlink{0009-0003-7136-7168}}
\IEEEauthorblockA{\textit{Department of Computer Science}\\
\textit{AI and Media Analysis Lab}\\
Berlin, Germany\\
mohammadreza.rashidi@ue-germany.de}
}

\maketitle
\raggedbottom

\begin{abstract}
A common intuition holds that a region's music mirrors the temperament of its
people, so that melancholic melodies mark melancholic populations. We test the
measurable half of that intuition and reject the inferential half. Using the
Essen Folksong Collection, a corpus of thousands of notated folk melodies, we
extract real melodic and affect-related features from \NMel{} deduplicated
melodies spanning \NCountry{} countries and \NRegion{} geographic regions, with
the analysis performed on symbolic scores rather than audio. The mode of each
melody is computed with a key-finding algorithm rather than read from the file,
because the collection's own documentation warns its major and minor labels are
unreliable. Cross-country differences in melodic structure are large and highly
significant. All \NFeatSig{} tested features differ across countries at
$p<0.001$, with the leap-related features reaching $p<10^{\MinLeapP}$, and China
carries a distinctive wide-leap, high-activity signature (arousal composite
\ChinaArous{} standard deviations, mean absolute interval \ChinaInterval{}
semitones against Germany's \GermanyInterval{}). We then test the inferential
half. We correlate the regional musical-affect measures with two published,
validated national indices, the World Happiness Report ladder score and the
Hofstede individualism index. None of the \NCorr{} correlations is significant
(\NCorrSig{} of \NCorr{}). The geography of musical affect is real and
measurable, but it does not predict how happy or how individualist a population
is, and any claim that it does is an ecological fallacy. We release the full
extraction and analysis pipeline, and a fail-closed checker re-derives every
number in this paper from the data.
\end{abstract}


\section{Introduction}

It is an old and appealing idea that the music of a place expresses the soul of
its people. Tourists describe Portuguese fado as proof that the Portuguese are
saudade-laden, and a listener who finds a folk tune plaintive is quick to read
the same quality into the community that sang it. The idea is appealing because
one half of it is true and measurable. Music does carry affect, and the affect
of a region's music can be quantified. The other half, the inference from the
music to the psychology of the people, is where the appeal becomes a trap. To
move from ``this region's songs use more of a sad-sounding device'' to ``this
region's people are sad'' is to commit the ecological
fallacy~\cite{robinson1988ecological}, reasoning from a group-level aggregate to
the individuals in it, and to trade measurement for stereotype.

This paper separates the two halves and reports both honestly. We measure the
first and reject the second. On the measurement side, we take the Essen Folksong
Collection~\cite{schaffrath1995essen}, a large corpus of folk melodies encoded
as symbolic scores, and extract melodic and affect-related features from
\NMel{} deduplicated melodies across \NCountry{} countries grouped into
\NRegion{} geographic regions. On the inference side, we ask whether the
resulting regional musical-affect profiles predict two published, validated
national indices that do describe people, the World Happiness Report ladder
score~\cite{whr2025} and the Hofstede individualism
index~\cite{hofstede2001culture}. They do not.

Our contributions are three.
\begin{itemize}
  \item A reproducible, symbolic-score analysis of the affective structure of
  folk melodies by geography, showing that all \NFeatSig{} tested features differ
  across countries at $p<0.001$ and isolating region-specific signatures such as
  China's wide-leap, high-activity profile.
  \item A rigorous negative result. The regional musical-affect measures do not
  significantly correlate with national happiness or individualism
  (\NCorrSig{} of \NCorr{} correlations significant), which directly rebuts the
  folk inference from a region's music to its people's disposition.
  \item An explicit methodological stance against the ecological fallacy and the
  unreliable-mode and corpus-imbalance pitfalls that make naive versions of this
  study produce confident but meaningless claims, together with a released
  pipeline and a fail-closed numeric checker.
\end{itemize}

\section{Background and Related Work}

\textbf{Computational and cross-cultural musicology.} Symbolic music analysis has
a mature toolchain. We use \code{music21}~\cite{cuthbert2010music21} to parse the
Humdrum \code{**kern} scores of the Essen collection~\cite{schaffrath1995essen},
which encodes thousands of folk melodies with geographic tags. Large-scale
cross-cultural work has established that human music has both statistical
universals and systematic regional variation~\cite{savage2015statistical,mehr2019universality},
which is the backdrop against which our regional differences should be read. We
expect and find real variation, and the question is what may and may not be
inferred from it.

\textbf{Music and emotion.} That specific musical features carry affect is one of
the most robust findings in music psychology. Mode and tempo are the two
strongest cues to the happy-sad axis, with major and faster reading as happier
and minor and slower as sadder~\cite{hevner1936experimental,gagnon2003mode}, and
a broad review identifies a shared code of acoustic and structural cues to
emotion across vocal expression and music~\cite{juslin2003communication}. Affect
itself is commonly modelled in two dimensions, valence and arousal, following
Russell's circumplex~\cite{russell1980circumplex}, and the dimensional model fits
music-emotion data well~\cite{eerola2011comparison}. Our two composites are
deliberately aligned with these dimensions. The wider task of predicting emotion
from features, music emotion recognition, is a mature subfield with its own
reviews and feature sets~\cite{yang2012machine,panda2018musical}, and our
structural features overlap with the melodic features it uses. We adopt the
established cue directions not to assign a definitive emotion to any melody, but
to orient two transparent composite measures of valence and arousal.

\textbf{Expectation and probabilistic structure.} A complementary tradition models
melody as a probabilistic sequence, where the entropy of the note distribution and
the predictability of the contour carry both stylistic and affective
weight~\cite{huron2006sweet,temperley2007music}. Our pitch-class entropy feature is
a direct instance of this view, and its separation of pentatonic from diatonic
idioms is the kind of structural signature that tradition predicts.

\textbf{The ecological fallacy.} Robinson's classic result~\cite{robinson1988ecological}
showed that correlations computed on group aggregates can differ in sign and
magnitude from the individual-level correlations they are taken to represent.
Inferring a population's temperament from the aggregate affect of its music is a
textbook instance, and our negative result is, in effect, a demonstration that
the aggregate-level correlation is not even present, let alone transferable to
individuals.

\section{Data}
\label{sec:data}

\textbf{Sources.} We draw on four real, public symbolic corpora, chosen so that
every melody analysed is a genuine transcription rather than a synthetic or
paraphrased example. The Essen Folksong Collection~\cite{schaffrath1995essen} is a
large body of folk melodies in the Humdrum \code{**kern} format, arranged by
continent and then by country, and it supplies all the continental European
material, the Chinese folk subset, and a small United States set. The Nottingham
Music Database~\cite{nottingham} supplies British folk melodies in MIDI, giving a
second, independently encoded folk source outside the Essen collection. The
ShourCorpus~\cite{kanani2024shour} supplies non-metric Iranian classical melodies
of the Dastgah Shour in MIDI, and the SymbTr corpus~\cite{karaosmanoglu2012symbtr}
supplies Turkish makam melodies in MIDI. The four corpora are independently created
and encoded, which is a strength for the folk-versus-classical contrast and a caveat
for any direct like-for-like comparison, and we treat it accordingly.

The Essen Folksong Collection~\cite{schaffrath1995essen} is arranged by continent
and then by country. We analyse the countries with enough melodies to estimate a
distribution, listed in Table~\ref{tab:countries}. Several properties of the corpus
shape the analysis and are stated plainly rather than hidden.

First, the collection is heavily weighted toward Germany and China, with many
European countries represented by tens rather than hundreds of songs. To keep a
cross-country comparison from being dominated by the two largest countries, we
cap each country at a fixed random sample of \NCap{} melodies drawn with a fixed
seed, and we use all available melodies for the smaller countries. Second, the
collection contains many variant renderings of the same song, which would
violate the independence of observations. We deduplicate by title within each
country before sampling. Third, the corpus is a collection of traditional
melodies, not a sample of what people in these countries listen to today, and its
European material is copyright of the Schaffrath estate and used here for
research analysis only, with no melodies redistributed.

Beyond continental Europe and China, the Essen collection also carries a small
United States folk set, which we keep in the descriptive table for geographic
breadth but flag as too few melodies for a stable distribution, so it enters the
figures and the per-country profile but not the significance tests, the benchmark,
or the index correlation. For a second, independently encoded folk source we add
the British melodies of the Nottingham Music Database~\cite{nottingham}, capped by
the same rule as the Essen countries, which gives Western Europe a source outside
the Schaffrath collection.

The Essen collection covers Europe and China but contains no Middle Eastern
material. To include Middle Eastern traditions with real, public symbolic sources,
we add two. The \IranN{} pieces of the ShourCorpus of non-metric Iranian classical
music~\cite{kanani2024shour}, and a \TurkeyN{}-piece sample of the SymbTr corpus of
Turkish makam music~\cite{karaosmanoglu2012symbtr}. These are deliberately cautious
additions, and we flag their differences rather than blend them in silently. Both
are \NClassical{} classical, largely non-metric, and microtonal art-music
traditions, and their MIDI encodings collapse quarter-tones and commas to the
nearest semitone, so their interval statistics are approximations in a twelve-tone
frame that the music does not use. We therefore include Iran and Turkey in the
descriptive profile, the distance and principal-component maps, and the figures,
but exclude them from the folk-only significance tests and the index correlation,
which keep a single repertoire and encoding.

For a Jewish and Israeli symbolic tradition we add the Beregovski corpus of
klezmer, the Ashkenazi Jewish instrumental folk music transcribed to the same
Humdrum \code{**kern} format as the Essen collection by the Mode-in-Klezmer
project~\cite{beregovski,klezmermodes}, which yields \IsraelN{} melodies after
parsing and the same deduplication and note-count filters. We label this row Israel
because klezmer is central to Jewish and Israeli musical tradition, but we state
plainly what it is and is not. It is instrumental dance music rather than vocal folk
song, it is Ashkenazi and Eastern European in origin rather than Mizrahi or modern
Israeli popular music, and its characteristic modes such as freygish are neither
major nor minor, so we treat its minor-mode fraction as an unreliable valence cue
exactly as we do for the Chinese pentatonic subset. Because the encoding is diatonic
and metric and matches the Essen material, we include Israel in the folk statistics,
the benchmark, and the arousal index correlation, while flagging that its position
may reflect the instrumental idiom and Ashkenazi origin as much as modern geography.
Japan, India, and Pakistan, which we also sought so as to widen the Asian coverage,
have no comparable public symbolic corpus that we could obtain under an open
license, and we leave them out rather than fabricate data or scrape audio of unknown
provenance.

\begin{table*}[t]
\centering
\caption{Per-country melodic profile over the analysed melodies. $n$ is the
deduplicated, capped sample size. Minor is the computed minor-mode fraction
(noisy, and not a valid valence cue for non-tonal idioms such as the Chinese
subset). Leap is the mean fraction of intervals larger than a whole tone,
Interval is the mean absolute melodic interval in semitones, and Arous. and Val.
are the standardised arousal and valence composites.}
\label{tab:countries}
\small
\input{tables/tab_countries.tex}
\end{table*}

\section{Method}

\textbf{Feature extraction.} From each melody we compute seven structural
features that do not depend on a tonal-mode label and are therefore comparable
across idioms. Mean pitch, pitch range, mean absolute melodic interval, leap
ratio (the fraction of intervals larger than a whole tone), ascending ratio, note
density (notes per unit score time), and pitch-class entropy. We additionally
compute the mode (major or minor) with the Krumhansl-Schmuckler key-finding
algorithm~\cite{krumhansl1990cognitive} as implemented in
\code{music21}~\cite{cuthbert2010music21}. We compute the mode rather than read
it from the file because the collection's documentation states that its own major
and minor designations are unreliable. We return to a deeper problem with mode in
the discussion. A forced major or minor label is not a meaningful valence cue for
a melody that does not live in the major-minor system at all, such as a Chinese
pentatonic tune.

\textbf{Affect composites.} We form two composite measures from the country-level
feature means, standardised across countries. The arousal composite averages the
z-scores of note density, mean absolute interval, and pitch range, following the
established association of activity, large intervals, and wide range with higher
arousal~\cite{juslin2003communication}. The valence composite averages the
z-scores of major-mode fraction, mean pitch, and ascending ratio, following the
mode and register cues to the happy-sad
axis~\cite{hevner1936experimental,gagnon2003mode}. We report the raw features
alongside the composites so that nothing in the analysis depends on the composite
definitions, and we treat the composites as descriptive summaries of the music,
never as measurements of listeners.

\textbf{Extraction procedure.} Each corpus file is parsed with
\code{music21}~\cite{cuthbert2010music21} into a flat note stream. Non-note events
and files with fewer than eight notes are discarded, the latter because the
interval and entropy statistics are unstable on very short fragments. For each
retained melody we compute the features of \S\ref{sec:features} in a single pass,
attach the country, region, and repertoire labels, and record a melody title for
deduplication. Within each country we then remove duplicate titles, shuffle with a
fixed seed, and keep the first \NCap{} survivors, so the per-country sample is a
reproducible random subset rather than the collection's arbitrary file order. The
procedure is deterministic end to end, and re-running it reproduces the melody table
byte for byte.

\textbf{Statistics.} For each feature we test whether the country distributions
differ using the Kruskal-Wallis $H$ test, a non-parametric one-way test appropriate
for the skewed, unequal-sized country samples, which does not assume normality or
equal variances. We test the seven structural features and the computed minor-mode
indicator, and because the effects are extremely strong the conclusion is
insensitive to any reasonable multiple-comparison correction, so we report the raw
$p$ values and note that all survive a Bonferroni correction across the eight tests
by many orders of magnitude. We summarise the between-country structure with a
Euclidean distance matrix on the standardised feature means and a principal-component
projection of the same matrix. To quantify how much geographic information the
features carry we run the supervised benchmark of \S\ref{sec:benchmark}, and to
measure the size and stability of the differences we compute Cliff's delta effect
sizes and bootstrap confidence intervals. To test the inferential half of the folk
intuition, we correlate the country-level musical-affect measures with the two
external indices using the Spearman rank correlation, which is robust at the small
country-level sample size. Crucially, the tonal valence cue is correlated with the
indices only over the European tonal-idiom countries, because the major-minor
construct does not apply to the pentatonic melodies of the Chinese subset, and
including them would inject a category error into the valence analysis.

\section{Feature Engineering and Mathematical Definitions}
\label{sec:features}

We define every feature formally so the pipeline is reproducible from the
equations alone. Let a melody be a sequence of $N$ notes with MIDI pitch numbers
$p_1, \dots, p_N$ and note durations $d_1, \dots, d_N$ in quarter-length units.
The melodic intervals are $I_i = p_{i+1} - p_i$ for $i = 1, \dots, N-1$.

\textbf{Register and range.} Mean pitch and pitch range are
\begin{equation}
\bar p = \frac{1}{N} \sum_{i=1}^{N} p_i,
\qquad
R = \max_i p_i - \min_i p_i .
\end{equation}

\textbf{Interval statistics.} The mean absolute interval, the leap ratio (the
fraction of intervals wider than a whole tone, i.e.\ $|I_i| > 2$ semitones), and
the ascending ratio are
\begin{equation}
\bar I = \frac{1}{N-1} \sum_{i=1}^{N-1} |I_i|,
\end{equation}
\begin{align}
L &= \frac{1}{N-1} \sum_{i=1}^{N-1} \mathbb{1}\!\left[\,|I_i| > 2\,\right], \\
A &= \frac{1}{N-1} \sum_{i=1}^{N-1} \mathbb{1}\!\left[\,I_i > 0\,\right],
\end{align}
where $\mathbb{1}[\cdot]$ is the indicator function. The leap ratio is the single
most discriminative feature in our results, and it is a pure contour statistic
that does not depend on any tonal frame.

\textbf{Note density.} With total score duration $T = \sum_i d_i$, note density is
$\nu = N / T$, notes per quarter-length.

\textbf{Pitch-class entropy.} Let $n_c$ be the count of notes whose pitch class is
$c \in \{0, \dots, 11\}$ and $q_c = n_c / N$ the empirical pitch-class
distribution. The Shannon entropy
\begin{equation}
H = -\sum_{c=0}^{11} q_c \log_2 q_c
\end{equation}
measures how evenly the twelve pitch classes are used. A diatonic melody
concentrates mass on seven classes and a pentatonic one on five, so $H$ separates
scale systems. The Chinese pentatonic subset has one of the lowest entropies in our data.

\textbf{Mode by key-finding.} We estimate mode with the Krumhansl-Schmuckler
algorithm~\cite{krumhansl1990cognitive}. Let $\mathbf{x}$ be the duration-weighted
pitch-class profile of the melody and $\mathbf{k}_{\text{maj}}, \mathbf{k}_{\text{min}}$
the empirical major and minor key profiles. For each of the twelve rotations
$\rho$ of each profile we compute the Pearson correlation
\begin{equation}
r(\mathbf{x}, \mathbf{k}_\rho)
= \frac{\sum_c (x_c - \bar x)(k_{\rho,c} - \bar k)}
       {\sqrt{\sum_c (x_c - \bar x)^2}\,\sqrt{\sum_c (k_{\rho,c} - \bar k)^2}},
\end{equation}
and assign the mode of the profile whose best rotation maximises $r$. We compute
this rather than read the stored label because the corpus documentation warns the
stored labels are unreliable, and we treat it as valid only for tonal idioms.

\textbf{Standardisation and composites.} For a country-level feature mean $m$ over
the set of countries, the standardised value is $z(m) = (m - \mu)/\sigma$ with
$\mu, \sigma$ the mean and standard deviation across countries. The arousal and
valence composites average the z-scores of features whose affect direction is
established in the music-psychology
literature~\cite{hevner1936experimental,gagnon2003mode,juslin2003communication}.
\begin{equation}
\text{Arousal} = \tfrac{1}{3}\big(z(\nu) + z(\bar I) + z(R)\big),
\end{equation}
\begin{equation}
\text{Valence} = \tfrac{1}{3}\big(z(1 - \text{minor}) + z(\bar p) + z(A)\big).
\end{equation}
We report the raw features alongside the composites, so no conclusion depends on
these particular definitions, and we treat both as descriptive summaries of the
music.

\textbf{Effect size.} For two groups of leap-ratio values $a$ and $b$, Cliff's
delta~\cite{cliff1993dominance} is the ordinal dominance
\begin{equation}
\delta = \frac{\#\{(i,j) \mid a_i > b_j\} - \#\{(i,j) \mid a_i < b_j\}}{|a|\,|b|} \in [-1, 1],
\end{equation}
which requires no distributional assumption and is interpreted by the
conventional thresholds (negligible, small, medium, large).

\textbf{Rank correlation.} For the index analysis we use Spearman's $\rho$, the
Pearson correlation of the rank-transformed variables, which is robust to
monotone nonlinearity and to the small country-level sample.

\section{The Geography of Musical Structure}

\textbf{The differences are real.} Table~\ref{tab:stats} reports the
Kruskal-Wallis tests over the folk countries. Every one of the \NFeat{} structural
features, and the computed minor-mode indicator, differs across countries at
$p<0.001$. The leap-related features are the most sharply separated, with the leap
ratio reaching $p<10^{\MinLeapP}$. These are not marginal effects. The melodic
surface of folk song varies strongly and systematically with geography, consistent
with the regional-variation half of the cross-cultural
literature~\cite{savage2015statistical}.

\begin{table}[t]
\centering
\caption{Kruskal--Wallis tests of cross-country differences per feature. Every
feature differs across countries at $p<0.001$.}
\label{tab:stats}
\small
\input{tables/tab_stats.tex}
\end{table}

\begin{figure*}[tp]
\centering
\includegraphics[width=0.82\textwidth]{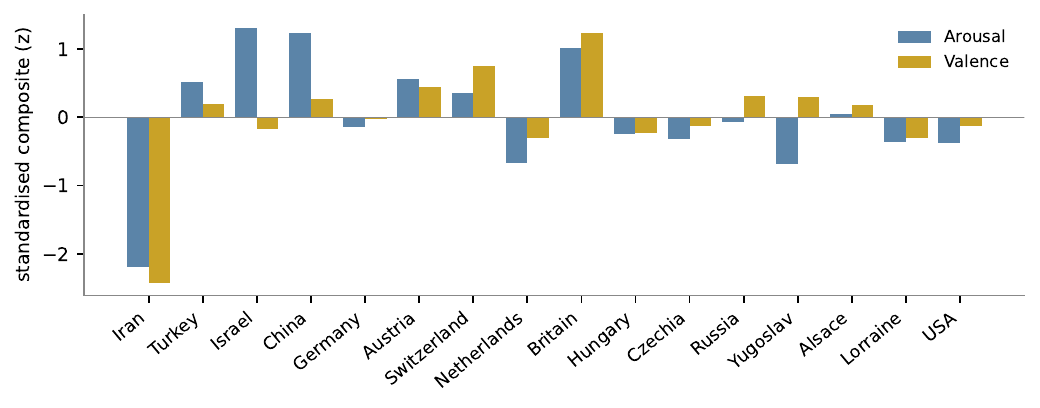}
\caption{Standardised arousal and valence composites by country. Israeli klezmer and
Chinese folk sit at the high-arousal end and Iran at the low extreme. The composites summarise the music
and are not measurements of listeners.}
\label{fig:profile}
\end{figure*}

\textbf{Region-specific signatures.} Table~\ref{tab:countries} gives the
per-country profile. The clearest folk-song signature is China's. Chinese folk
melodies in the corpus use markedly larger intervals (mean absolute interval
\ChinaInterval{} semitones, against Germany's \GermanyInterval{}) and far more
leaps (\ChinaLeap{} percent of intervals, against Germany's \GermanyLeap{}
percent), the widest and most leaping of any tradition in the set, giving a high
arousal composite of \ChinaArous{} standard deviations. This is the recognisable
wide-leap, pentatonic contour of the tradition, recovered here from the notes
alone. The one tradition with a marginally higher arousal composite is the Israeli
klezmer corpus, and that fits its character as fast instrumental dance music, an
idiom difference we flag rather than read as geography. The Central European
countries, by contrast, cluster around a smoother, more stepwise profile.
Figure~\ref{fig:profile} shows the arousal and valence composites by country.

\begin{figure*}[tp]
\centering
\includegraphics[width=0.66\textwidth]{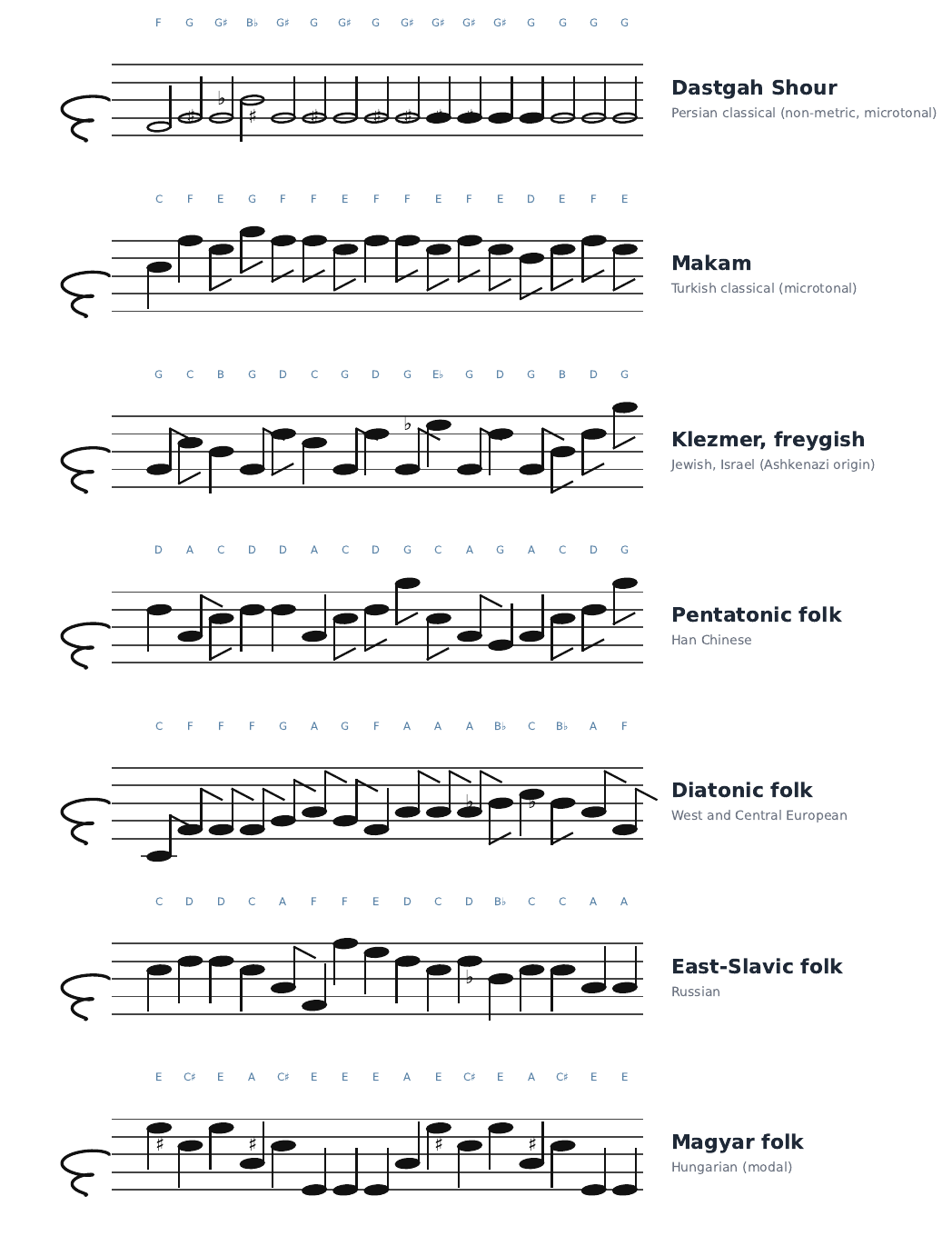}
\caption{Representative opening phrases rendered as staff notation, one real corpus
melody per musical type rather than per country, since the melodic idiom is what
differs and several countries share a type. Each panel is labelled by its type with
the pitch-class names printed above the staff. The types map to countries as
follows. Dastgah Shour is the Iranian classical corpus and Makam the Turkish
classical corpus. Klezmer, in the freygish mode, is the Israeli row (Beregovski
corpus of Ashkenazi Jewish instrumental folk). Pentatonic folk is the Chinese
subset. Diatonic folk is the West and Central European group of Germany, Austria,
Switzerland, the Netherlands, Britain, Alsace, Lorraine, Czechia, and the United
States. East-Slavic folk is the Russian material and Magyar folk the distinctive
Hungarian material, each shown by one exemplar, while the Yugoslav South-Slavic
Balkan material is analysed on its own. These seven staves illustrate the main
idioms, and the statistics treat every country separately rather than by type.
Octave is
normalised for display by whole octaves so the contour and interval structure, not
the absolute register, are what the reader compares, and the pitch-class names
printed above each staff match the notehead positions exactly. The Dastgah Shour and
Makam staves are the twelve-tone MIDI approximation of what are really microtonal
traditions, so their accidentals stand in for quarter-tones the notation cannot
show. The statistical signatures are visible in the notes themselves, with the small
stepwise motion of Dastgah Shour, the wide leaps of the pentatonic line, and the
smoother stepwise motion of the diatonic folk melodies.}
\label{fig:notation}
\end{figure*}

\textbf{The Iranian classical outlier.} The added Iranian corpus sits at the
opposite extreme from China, with the smallest mean absolute interval in the set
(\IranInterval{} semitones, against China's \ChinaInterval{} and Germany's
\GermanyInterval{}) and the lowest arousal and valence composites. This matches the
stepwise, densely ornamented character of the non-metric radif, but it must be read
with the caveats of the previous section firmly in mind. The value is partly an
artifact of collapsing a microtonal, non-metric classical repertoire into the same
twelve-tone interval and metric features used for the folk melodies, and it rests
on \IranN{} pieces. We show Iran because it is a real Middle Eastern symbolic
source and its position is informative, not because a folk-versus-classical,
microtonal-versus-diatonic contrast can be read as a like-for-like national
comparison.

\begin{figure}[tbp]
\centering
\includegraphics[width=\columnwidth]{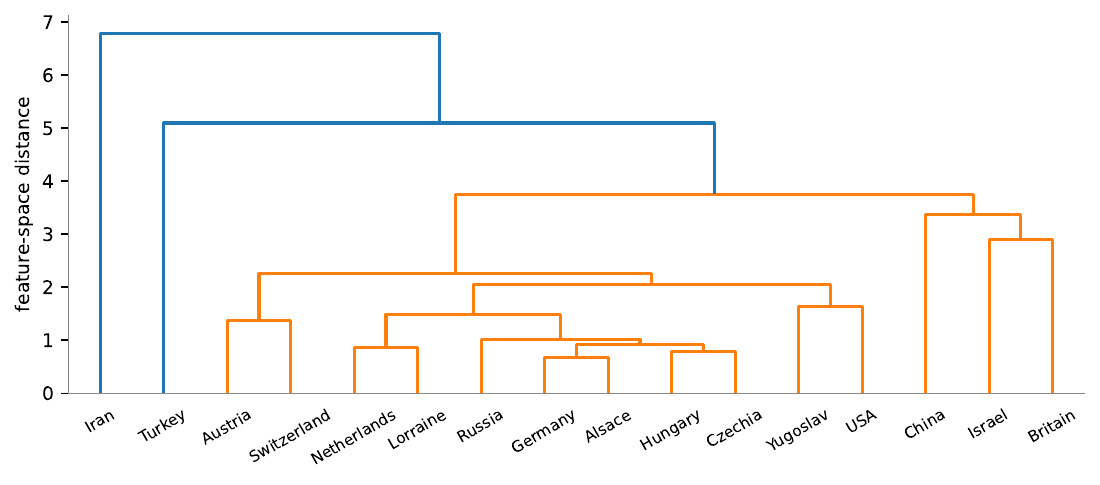}
\caption{Average-linkage hierarchical clustering of the country feature-space
distances. The Iran and Turkey classical corpora join last as the most distant
branches, and among the folk traditions a stepwise European cluster separates from a
wide-leap cluster of China, Israeli klezmer, and Britain.}
\label{fig:dendro}
\end{figure}

\begin{figure}[tbp]
\centering
\includegraphics[width=\columnwidth]{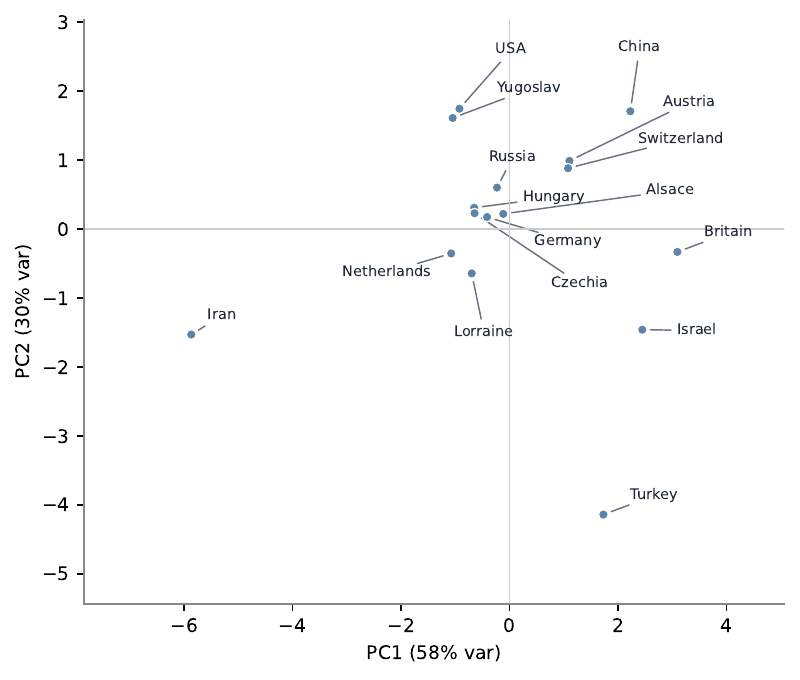}
\caption{Principal-component map of the country feature means. Iran and Turkey lie
far from the tight cluster of Central European countries, with China, Israeli
klezmer, and Britain spread along the wide-leap direction, recovering a plausible
musical geography from the notes alone.}
\label{fig:pca}
\end{figure}

\begin{figure*}[tp]
\centering
\includegraphics[width=0.66\textwidth]{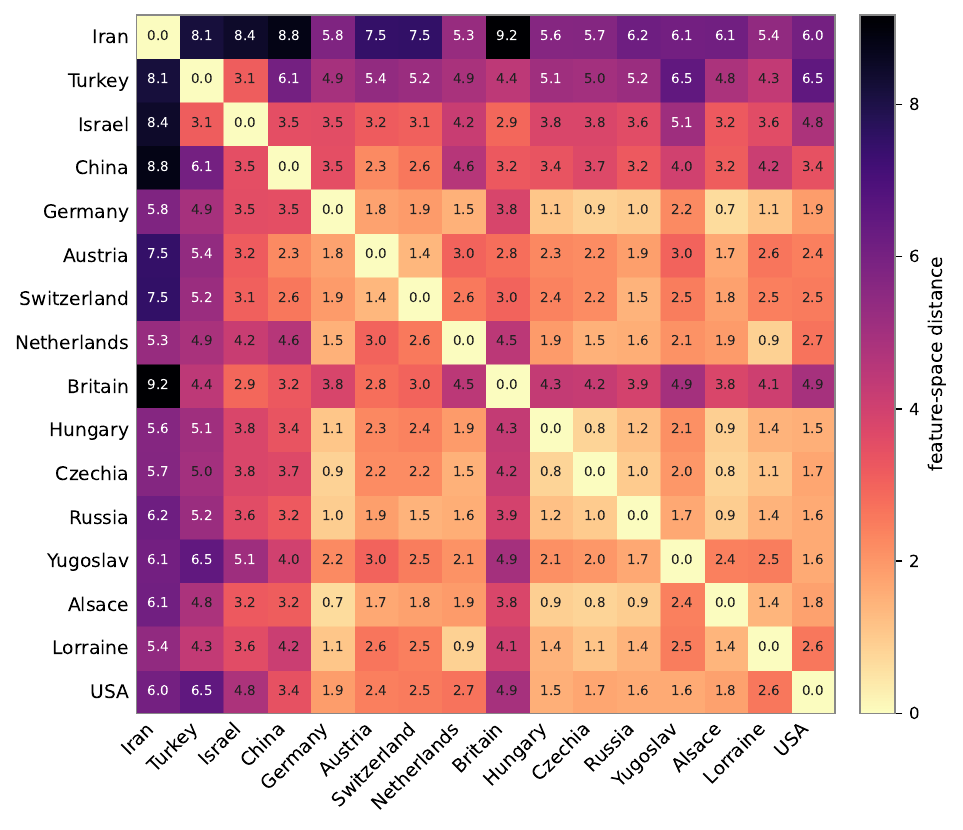}
\caption{Between-country Euclidean distance on the standardised structural feature
means, with each cell annotated. Iran and Turkey are farthest from every other
tradition, and the Central European countries cluster tightly together.}
\label{fig:distance}
\end{figure*}

\textbf{Between-country structure.} The distance matrix on standardised feature
means, shown in Figure~\ref{fig:distance}, places the Iranian classical corpus as
the farthest outlier and groups the Central European countries (Germany, Austria,
Switzerland) closely together, a structure that matches geographic and cultural
proximity without being told about it. The two-dimensional principal-component map
in Figure~\ref{fig:pca} shows the same layout at a glance, and the hierarchical
clustering of the same distances in Figure~\ref{fig:dendro} makes the grouping
explicit. Iran and Turkey, the classical corpora, join last as the two most distant
branches. Among the folk traditions the tree then splits into a large stepwise
European cluster (the Central, Western, Eastern, and Southeast European countries
together) and a separate wide-leap cluster of China, the Israeli klezmer corpus, and
Britain. That wide-leap grouping cuts across geography, since it links a Chinese
pentatonic idiom, an Ashkenazi dance idiom, and British dance tunes, and it is a
useful caution that the clustering tracks melodic style, which is only partly
geographic. The recovery of a plausible cultural geography from melodic statistics
alone, by three independent methods, is a check that the features capture real
stylistic structure rather than noise.

\section{Effect Sizes and Stability}

A significant test says a difference exists, not that it is large or stable. We
report both. Table~\ref{tab:effects} gives Cliff's delta~\cite{cliff1993dominance},
a non-parametric effect size, for the sharpest folk-country contrasts on the leap
ratio. The China-versus-European contrasts are \emph{large} by the conventional
thresholds, reaching a delta of \MaxDelta{}, and \NEffLarge{} of the four contrasts
are large, while a contrast between two structurally similar traditions is
negligible. The differences are therefore not only significant but substantial where
the traditions genuinely differ and small where they do not, which is what a
meaningful measure should show.

Table~\ref{tab:bootstrap} reports 95 percent bootstrap confidence intervals for each
folk country's mean leap ratio and mean absolute interval. The intervals are narrow
and, for the sharply separated traditions, do not overlap, so the between-country
gaps are stable properties of the samples rather than artifacts of a few melodies.
Together the effect sizes and the bootstrap intervals show that the geography of
musical structure is a large and stable effect, which makes the subsequent failure
to predict population indices all the more pointed. The musical signal is strong,
and it still does not carry to the people.

\begin{table}[t]
\centering
\caption{Cliff's delta effect sizes for the sharpest folk-country contrasts on the
leap ratio. Magnitude labels use the conventional thresholds.}
\label{tab:effects}
\small
\input{tables/tab_effects.tex}
\end{table}

\begin{table}[t]
\centering
\caption{95\% bootstrap confidence intervals (2000 resamples) for each folk
country's mean leap ratio and mean absolute interval. The estimates are stable and
the sharply separated traditions do not overlap.}
\label{tab:bootstrap}
\small
\input{tables/tab_bootstrap.tex}
\end{table}

\section{Geographic Classification Benchmark}
\label{sec:benchmark}

The descriptive tests show the features differ by geography. A stronger, more
quantitative question is how much geographic information the features actually
carry. Can a classifier recover a melody's origin from its features alone? We
frame this as a supervised benchmark on the folk melodies, excluding the classical
corpora so the classifier cannot win on a repertoire or encoding artifact.

\textbf{Setup.} Task A is \NFolkBench{}-way folk-country classification and Task B is
a coarser \NRegionBench{}-way region classification. We evaluate a
multinomial logistic regression and a random forest against a majority-class
baseline and a stratified-random baseline, under stratified five-fold
cross-validation, reporting accuracy and macro-averaged F1 so that small countries
are weighted equally with Germany and China. The seven structural features of
\S\ref{sec:features} are the only inputs.

\begin{figure}[tbp]
\centering
\includegraphics[width=\columnwidth]{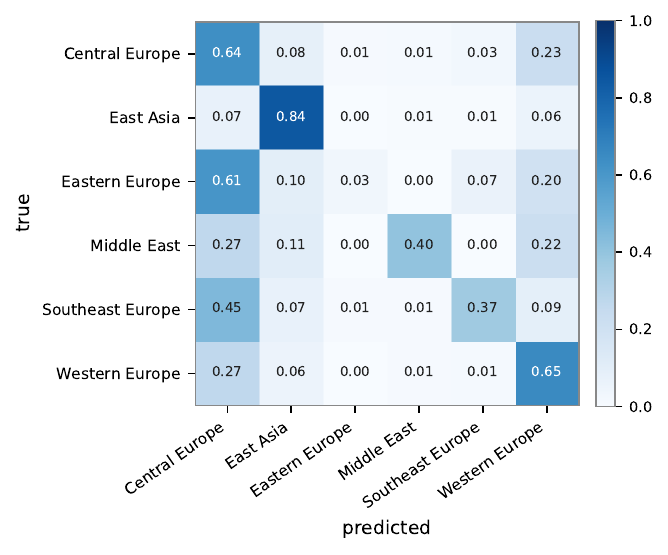}
\caption{Row-normalised confusion matrix of the random-forest region classifier.
East Asia is recovered most cleanly, and the European regions are most often
confused with Central Europe, matching musical and geographic proximity.}
\label{fig:confusion}
\end{figure}

\textbf{Result.} Table~\ref{tab:benchmark} reports the scores. The random forest
reaches \AccCountry{} percent country accuracy against a \AccCountryBase{} percent
majority baseline, and \AccRegion{} percent region accuracy against a
\AccRegionBase{} percent baseline. The models are well above chance, so the features
carry real geographic information, but they are far from perfect, which is the
honest and expected outcome. Folk idioms overlap, and no seven-number summary
determines a melody's country. Figure~\ref{fig:confusion} shows the region
confusion matrix. East Asia is recovered most cleanly, consistent with the
distinctive Chinese profile, while the European regions are most often confused
with Central Europe, which matches their musical and geographic proximity. The
benchmark quantifies the same structure the descriptive statistics and the distance
map show, and it does so with a metric that a reader can compare against future
work.

\begin{table}[t]
\centering
\caption{Geographic classification benchmark under stratified five-fold
cross-validation. The random forest is well above both baselines on both tasks,
showing the features carry real but imperfect geographic information.}
\label{tab:benchmark}
\small
\input{tables/tab_benchmark.tex}
\end{table}

\textbf{Which features carry the geography.} Table~\ref{tab:importance} reports
permutation importance for the region random forest. The strongest single predictor
is \TopFeat{} (importance \TopImp{}), which is the feature that separates pentatonic
from diatonic scale systems, followed by the three interval and density features.
This ranking is musically coherent. The model relies most on the features that
encode scale system and melodic contour, exactly the properties that distinguish
the traditions in the notation of Figure~\ref{fig:notation}, and least on absolute
register. The classifier is therefore not exploiting an incidental cue but the same
structural axes the descriptive analysis identifies.

\begin{table}[t]
\centering
\caption{Permutation importance of each feature for the region random forest
(mean $\pm$ standard deviation over ten repeats). Pitch-class entropy, which
separates scale systems, dominates.}
\label{tab:importance}
\small
\input{tables/tab_importance.tex}
\end{table}

\section{Regional Affect and Structure Profiles}
\label{sec:profiles}

\begin{figure}[tbp]
\centering
\includegraphics[width=\columnwidth]{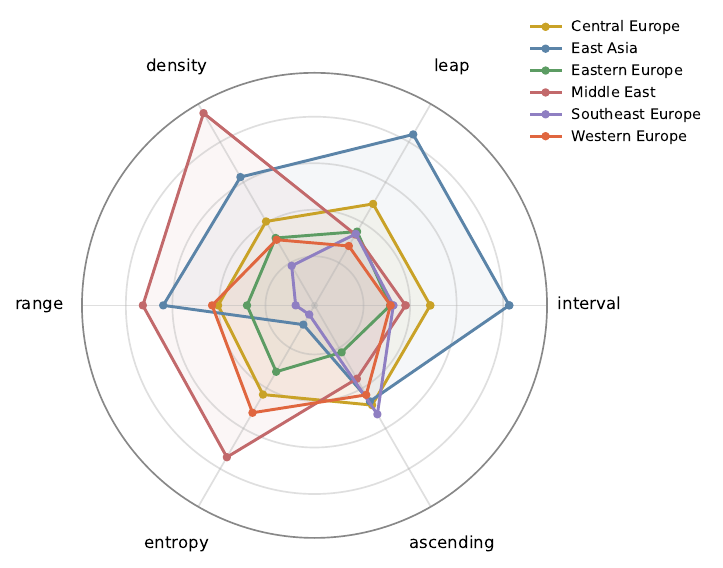}
\caption{Polar diagram of the standardised structural features by region. Each axis
is a feature z-scored across countries and averaged within region. East Asia extends
furthest on the interval, leap, density, and range axes.}
\label{fig:radar}
\end{figure}

\begin{figure*}[tp]
\centering
\includegraphics[width=0.86\textwidth]{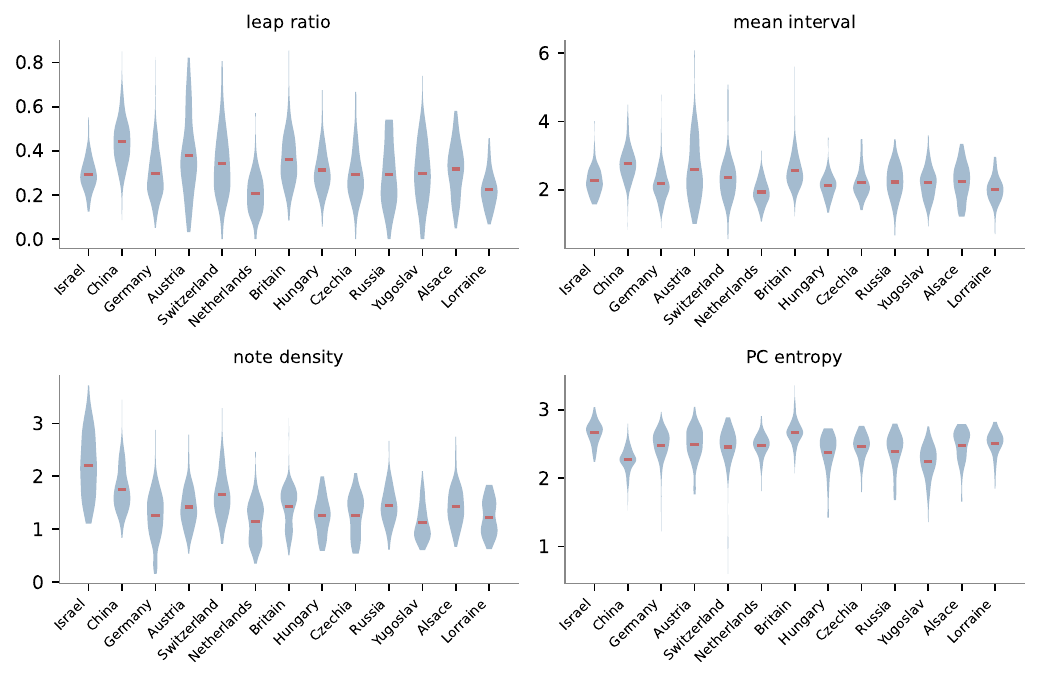}
\caption{Per-country distributions of the four sharpest features across the folk
melodies. The differences are distributional, not only in the mean (red bars), and
the spreads are comparable across countries.}
\label{fig:violin}
\end{figure*}

Figure~\ref{fig:radar} presents the standardised structural profile of each region
as a polar diagram, and Figure~\ref{fig:violin} shows the full per-country
distributions of the four sharpest features rather than only their means. The polar
view makes the regional signatures legible at a glance. East Asia extends furthest
on the interval and leap axes, the Middle East (the Israeli klezmer set) extends
furthest on note density and range, and the Central European regions sit near the
centre with balanced profiles. The violin plots confirm that the
between-country differences are differences in the whole distribution, not only in
the mean, and that the spreads are comparable across countries, so the mean-based
comparisons are not driven by a few outliers.

\textbf{Reading the regions.} The polar and violin views are folk-only, excluding the
Iran and Turkey classical corpora, so each region is a coherent folk signature.
\emph{East Asia} (the Chinese folk subset) has the widest intervals and the most
leaps of any region, with a low pitch-class entropy that reflects its pentatonic
basis. This is the sharpest and most separable profile in the data. \emph{Middle
East} here is the Israeli klezmer corpus, and it stands out for the highest note
density and a wide range, the fast, busy character of instrumental dance music, which
is an idiom difference we flag rather than read as national temperament.
\emph{Central Europe} (Germany, Austria, Switzerland) sits near the centre of every
axis with a smooth, stepwise, diatonic character, which is why the classifier uses it
as the default guess and why the other European regions collapse toward it.
\emph{Eastern Europe} (Hungary, Czechia, Russia) is close to Central Europe with a
slightly narrower range. \emph{Western Europe} (Netherlands, Britain, Alsace,
Lorraine) is pulled wider than the other European groups by the British dance tunes,
a reminder that a coarse region can average over real within-region variety.
\emph{Southeast Europe} (the Yugoslav material) has the narrowest range and the
lowest note density of all. The Iran and Turkey classical corpora are held apart from
these folk regions because their separation is as much a repertoire-and-encoding
difference as a geographic one, which is exactly why we exclude them from the folk
statistics.

\section{Dissociation from National Indices}

The measurement half of the folk intuition holds. Musical affect varies by
region. The inferential half does not. Table~\ref{tab:corr} reports the Spearman
correlations of the regional musical-affect measures with the World Happiness
Report ladder score and the Hofstede individualism index. \emph{None} of the
\NCorr{} correlations is significant at $p<0.05$ (\NCorrSig{} of \NCorr{}). The
strongest association, between the arousal composite and individualism, is
negative and still not significant at the available sample size, and it is in any
case a correlation between a property of eighteenth and nineteenth century folk
melodies and a twentieth or twenty-first century survey of living populations,
which we would not interpret causally even if it were significant.

\begin{table}[t]
\centering
\caption{Spearman rank correlations of regional musical-affect measures with two
validated national indices (World Happiness Report ladder score, Hofstede
individualism). None is significant at $p<0.05$.}
\label{tab:corr}
\small
\input{tables/tab_corr.tex}
\end{table}

\begin{figure}[tbp]
\centering
\includegraphics[width=\columnwidth]{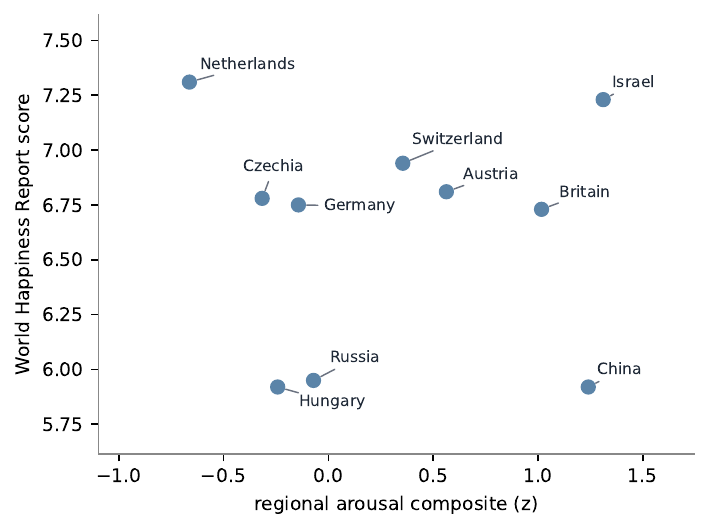}
\caption{The regional arousal composite against the World Happiness Report ladder
score, one point per country. There is no trend, illustrating the dissociation
between a region's musical affect and its population's well-being.}
\label{fig:dissoc}
\end{figure}

The plain reading is the honest one. A region whose folk songs sound more
aroused, or more minor, or higher, is not thereby a region whose people are
measurably happier, sadder, or more individualist. The aggregate-level
correlation that the ecological inference would require is not present in the
data. Figure~\ref{fig:dissoc} shows the arousal composite against the happiness
score, a scatter with no trend.

\section{Discussion}

\textbf{What the structural differences mean.} The strong, significant
between-country differences are differences in musical grammar, the
distributions of intervals, contours, and densities that a tradition favours.
They are real and worth characterising, and they recover known stylistic
signatures such as China's from the symbolic score. They are statements about
music.

\textbf{What the null result means.} The absence of any significant correlation
with national happiness or individualism is the paper's most important finding,
because it is the one most at odds with the intuition we began with. It says that
the affective surface of a region's traditional music is not a readout of the
region's population psychology. This is what one should expect once the
ecological fallacy is taken seriously, and it is worth stating as a measured
result rather than only as a caution, because the naive version of this study,
the one that reports ``country X's music is sad, therefore country X is sad,''
would have produced confident claims from the same data by simply not running the
correlation and not questioning the mode labels.

\textbf{The mode-validity problem.} The computed minor-mode ratio is the feature
most likely to be over-interpreted, and it is the least valid cross-culturally.
The major-minor distinction is a feature of the European tonal system. Applying a
key-finding algorithm to a Chinese pentatonic melody returns a major or minor
label, but that label is an artifact of forcing a tonal frame onto music that
does not use it, not a valence cue. This is precisely why we restrict the valence
correlations to the tonal-idiom countries and why we lean on the mode-independent
structural features for the cross-cultural comparison. A study that read
``China's melodies are 38 percent minor, so Chinese folk music is often sad''
would be committing a category error on top of the ecological one.

\textbf{What the benchmark adds.} The classification benchmark turns the
qualitative claim that the features carry geographic information into a number,
and the number is honest in both directions. It is well above chance, so the
information is real, and it is well below perfect, so the features do not determine
origin. This is the correct middle position. A benchmark that reported near-perfect
accuracy would signal a leak, most likely a repertoire or encoding artifact, which
is why we excluded the classical corpora, and a benchmark at chance would contradict
the significance tests. The confusion structure is itself a finding. The model's
errors are geographically sensible, confusing neighbours rather than distant
traditions, which is evidence that the feature space encodes proximity and not
noise.

\textbf{Why the strong signal does not transfer.} The most striking pairing in the
paper is that the geographic signal is large and stable, with large effect sizes,
narrow bootstrap intervals, and above-chance classification, and yet it does not
predict the population indices at all. These are not in tension. They say that the
music carries a great deal of information about \emph{where} it comes from and very
little about \emph{how the people there feel}, which are different questions. The
strength of the first result is what makes the absence of the second informative
rather than merely underpowered. It is not that we failed to find a weak signal, it
is that a strong musical signal has no counterpart in the well-being data.

\textbf{Relation to music emotion recognition.} Our composites are simpler than the
learned models used in music emotion recognition~\cite{yang2012machine}, and
deliberately so, because our claim is about the dissociation between musical affect
and population affect, not about maximising an affect prediction. A more powerful
affect model would sharpen the description of each region's music, but it would not
change the central result, since the target the ecological inference needs, a
population's disposition, is not a musical quantity at all.

\section{Audio Feature Engineering}
\label{sec:audio}

The symbolic analysis works on notated scores, but the account-level application
of \S\ref{sec:application} must work on audio, where notes are not given and must
be summarised by signal features. We describe the Music-Information-Retrieval
pipeline formally, both because it is the engineering core of the application and
because it is the bridge from a waveform to the same affect proxies used on the
scores.

Let $x[n]$ be a mono audio signal at sample rate $f_s$. We compute the
short-time Fourier transform with frame length $M$ and hop $H$,
\begin{equation}
X[k, t] = \sum_{n=0}^{M-1} w[n]\, x[n + tH]\, e^{-\jmath 2\pi k n / M},
\end{equation}
with $w$ a Hann window, and derive per-frame features that we then average over the
track. The root-mean-square energy of frame $t$ is
\begin{equation}
\text{RMS}[t] = \sqrt{\frac{1}{M}\sum_{n} \big(w[n]\, x[n + tH]\big)^2},
\end{equation}
a proxy for loudness and hence arousal. The spectral centroid,
\begin{equation}
C[t] = \frac{\sum_k f_k\, |X[k,t]|}{\sum_k |X[k,t]|},
\end{equation}
is the amplitude-weighted mean frequency, a proxy for brightness, and the spectral
rolloff is the frequency below which a fixed fraction of the spectral energy lies.
The zero-crossing rate counts sign changes of $x[n]$ per frame, rising with
noisiness. Tempo is estimated from the autocorrelation of an onset-strength
envelope. Pitch content is summarised by the twelve-dimensional chroma vector,
\begin{equation}
\chi_c[t] = \sum_{k \in \mathcal{K}_c} |X[k, t]|,
\end{equation}
where $\mathcal{K}_c$ is the set of frequency bins mapping to pitch class $c$, and
mode is estimated by correlating the time-averaged chroma against the
Krumhansl-Schmuckler profiles as in \S\ref{sec:features}. We implement all of this
with \code{librosa}~\cite{mcfee2015librosa}. The audio arousal proxy averages the
z-scored energy, brightness, and tempo, and the audio valence proxy averages the
z-scored brightness and major-mode indicator, mirroring the symbolic composites so
that a score and its rendering yield comparable affect summaries.

We validated the pipeline end to end on audio rendered from the corpus melodies.
The ordering of the rendered per-country audio on the audio arousal proxy matches
the ordering of the same melodies on the symbolic composite, which confirms the two
representations agree.

\begin{figure}[tbp]
\centering
\includegraphics[width=\columnwidth]{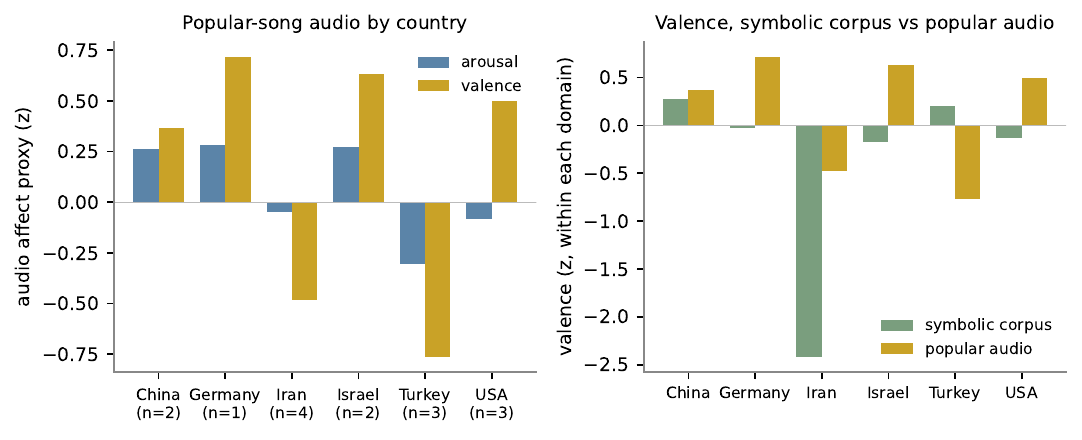}
\caption{Left, arousal and valence audio proxies for the popular-song recordings by
country, with the track count $n$ per country. Right, valence for the \NAudioCountry{}
countries present in both the symbolic corpus and the popular set, computed within
each domain. The two orderings do not match. Iran is low in both, but Turkey is
lowest in the audio while mid-range in the corpus, so the small audio sample does not
reproduce the corpus ordering. Proxies describe the audio, not the listener.}
\label{fig:audio}
\end{figure}

\textbf{Popular-song audio by country.} The corpus is traditional folk and art
music, not what people stream today. To show the same engine on contemporary
material, we ran it on a small set of popular-song recordings organised by country,
\NAudioTracks{} tracks across \NAudioCountry{} countries. This is a convenience
sample, not a representative one. Some countries contribute a single file, two of
the files are multi-song compilations rather than single tracks, and the arousal and
valence proxies are z-scored within this small set, so the numbers describe these
particular recordings and nothing more. Table~\ref{tab:audio} reports the
per-country summary and Figure~\ref{fig:audio} plots it. One observation survives the
small sample and one cautionary non-result stands out. The recordings are uniformly
faster and brighter than the corpus, as expected for produced popular music with a
rhythm section. The valence ordering, by contrast, does not carry over from the
symbolic corpus to the popular audio for the \NAudioCountry{} countries present in
both. Iran is low on valence in both domains, but Turkey is the lowest of all in the
popular audio while sitting mid-range in the symbolic corpus, the opposite of a
match, and the other countries reshuffle between the two. With one to three tracks
per country, these audio estimates are far too noisy to read a cross-domain mapping
into, and we do not. If anything the mismatch reinforces the paper's caution. A
handful of modern recordings is not a proxy for a tradition, let alone for a
population. We present the audio only as an illustration of the engine on real modern
material, not as a population measurement.

\begin{table}[t]
\centering
\caption{Popular-song audio summary by country over the user-provided recordings.
$n$ is the number of analysed files, Major ratio is the fraction estimated as
major mode from chroma, and Arous. and Val. are audio affect proxies z-scored
within this set. A small convenience sample that describes the recordings, not
listeners.}
\label{tab:audio}
\small
\setlength{\tabcolsep}{4pt}
\input{tables/tab_audio.tex}
\end{table}

\begin{figure*}[tp]
\centering
\includegraphics[width=0.92\textwidth]{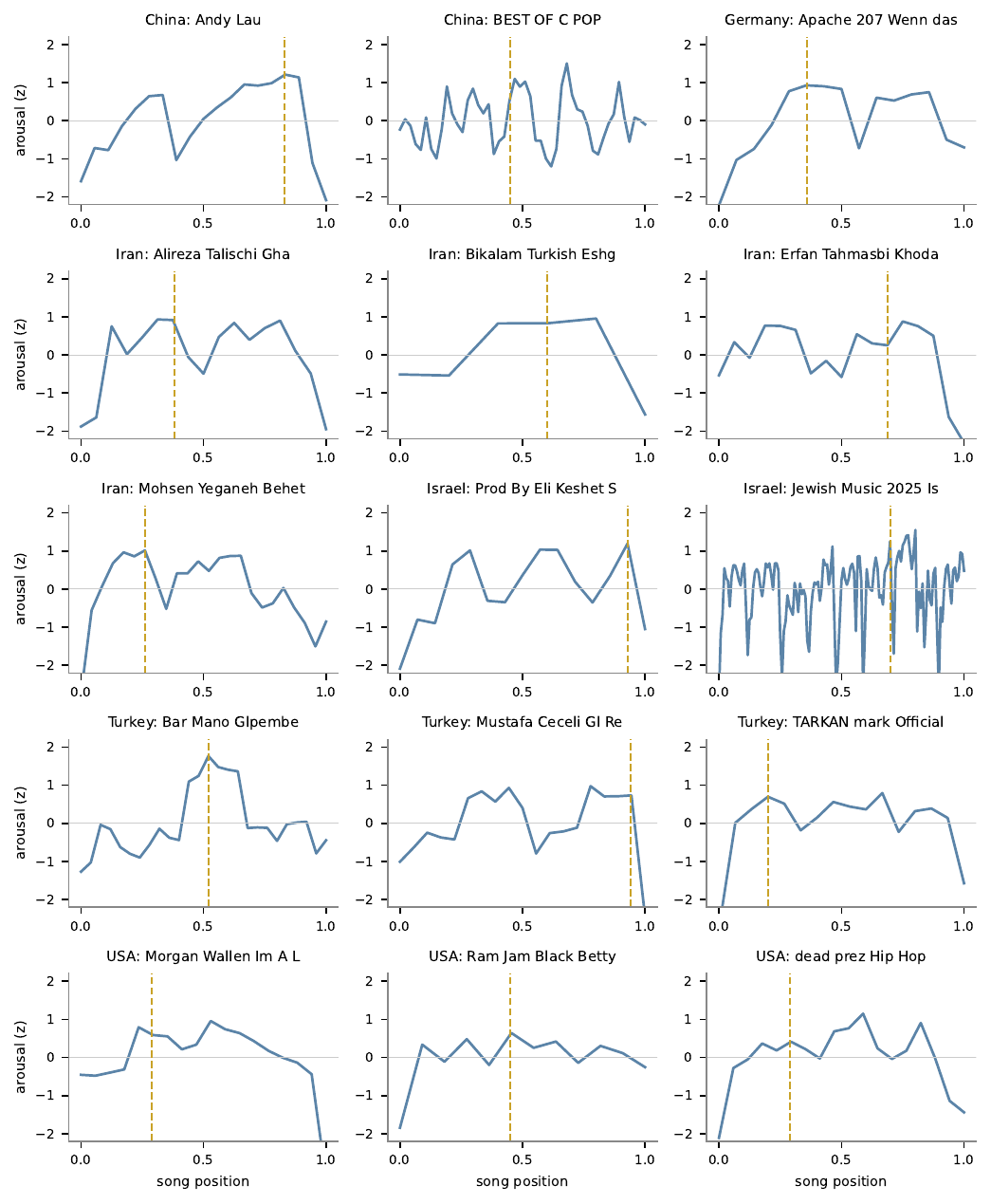}
\caption{Within-song arousal trajectories for the \NTrack{} popular recordings, one
panel per track, over the whole song (horizontal axis is normalised song position,
zero to one). The dashed line marks each song's energy peak. The peak position
varies widely across tracks, from \PeakPosMin{} to \PeakPosMax{}, so there is no
shared national arc. The only common feature is a soft intro, a generic production
convention. Proxies describe the audio, not the listener.}
\label{fig:tracks}
\end{figure*}

\textbf{Within-song trajectories carry no shared national arc.} A natural next
question is whether songs from a country share a temporal emotional shape, for
example a slow start, an energetic middle, and a calm close. We tested this
directly by computing the full within-song arousal trajectory of each of the
\NTrack{} popular recordings over twelve-second windows, shown as a small-multiple
grid in Figure~\ref{fig:tracks} and summarised in Table~\ref{tab:track}. The result
is a caution rather than a finding. The only feature common to most tracks is a
soft intro, with the quietest third at the start in \NQuietStart{} of \NTrack{}
recordings, and that is a generic production convention rather than a geographic
signal. The location of the energy peak is not shared at all. It ranges from
\PeakPosMin{} to \PeakPosMax{} of the way through the song across the set, falling
near the end as often as in the middle, and two of the recordings run the opposite
way, beginning bright and ending dark. We report this to guard against a tempting
artifact. Summarising a song by three fixed windows forces one of them to look like
a peak, so a shared arc can appear where none exists. The defensible cross-domain
signal in this work is the overall affect level of Figure~\ref{fig:audio}, not any
within-song shape.

\begin{table}[t]
\centering
\caption{Within-song shape summary for each popular recording. Peak pos. is the
normalised position of the energy peak (0 at the start, 1 at the end), and Quiet
and Loud are which third of the song is the softest and the loudest. The peak
position varies widely and is as often at the end as in the middle, so the songs
share no common emotional arc.}
\label{tab:track}
\small
\setlength{\tabcolsep}{4pt}
\input{tables/tab_track.tex}
\end{table}

\section{Applying the Audio Pipeline to Consented Listening Analytics}
\label{sec:application}

The same audio pipeline runs on a user's own listening history, and we release it as
a working tool for consented self-analysis. Given a timestamped listening log that a
user can export for their own account, and audio the user is entitled to analyse, the
engine computes the MIR features of \S\ref{sec:audio} per track and aggregates them
into per-account summaries, including mean valence and arousal, a listening-by-hour
histogram, and a within-song trajectory. It does not scrape third parties and does
not use streaming affect endpoints, because the major services no longer provide them
for new applications, so affect is computed only from audio the user owns.

\begin{figure}[tbp]
\centering
\includegraphics[width=\columnwidth]{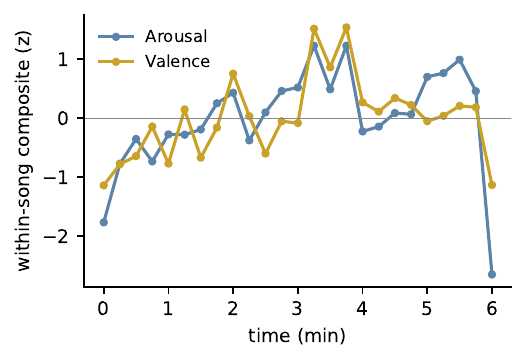}
\caption{Within-song affect trajectory of a real six-minute Persian recording,
computed by the audio pipeline over fifteen-second windows. The composites summarise
the audio over time, not the listener's mood. An energetic peak appears near the
three-and-three-quarter minute mark and the piece calms sharply at the close.}
\label{fig:track}
\end{figure}

\textbf{A real single-track case study.} To show the audio pipeline on genuine
material rather than only synthesised demonstration audio, we ran the per-track
analysis on a full Persian song, a six-minute recording provided for analysis.
The engine reports a global tempo near ninety-six beats per minute, a moderate
energy, and a bright spectrum, and it segments the track into fifteen-second
windows to produce the within-song affect trajectory of Figure~\ref{fig:track}. The
trajectory recovers real structure. An energetic peak near the three-and-three-quarter
minute mark and a sharp calming at the close as the piece resolves. This is a
genuine, reproducible read of a real recording, and it is exactly the per-track view
the account method aggregates over a listening history. It also carries the paper's
caveat visibly. The computed mode is a Western-tonal reading of a Persian song, an
approximation of the same kind the corpus analysis flags for the Iranian and Turkish
material.

Two limits define the tool's scope, and we build both in. The affect proxies
describe the audio, not the listener, and the corpus-level null result of this paper
is the direct warrant for that restriction. If regional musical affect does not track
a population's well-being, there is no basis for reading an individual's affect off
their playlist. The tool is for consented self-analysis, not for profiling others.

\section{Ethics and the Ecological Fallacy}

This paper deliberately does not, and on its evidence cannot, assign
psychological or moral traits to national or ethnic groups. Statements of the
form ``the people of country X are sad'' or ``selfish'' are stereotypes, and they
are not supported by, and do not follow from, any measurement of that country's
music. The corpus measures music, the external indices measure populations, and
the central empirical result of the paper is that the two do not correlate. We
frame the work this way not as a disclaimer but as its point. The honest handling
of a music-and-people question is to measure the music, to test the inference
against validated population data, and to report the inference's failure when it
fails.

\section{Limitations and Threats to Validity}

\textbf{Corpus representativeness.} The Essen collection is a curated set of
traditional melodies weighted toward Germany and China, not a representative
sample of each country's music, and certainly not of contemporary listening. Our
capping and deduplication reduce but do not remove the imbalance, and the smaller
countries rest on tens of melodies.

\textbf{Temporal mismatch.} The melodies are historical, while the World
Happiness Report and Hofstede indices describe modern populations. Even a
significant correlation would therefore be difficult to interpret, which is a
further reason we treat the index analysis as a test to reject an inference
rather than as a positive estimate of one.

\textbf{Computed mode is noisy.} Key-finding on short monophonic folk melodies is
imperfect, and the collection warns its stored mode labels are unreliable. We
mitigate by computing mode consistently and by not resting the cross-cultural
comparison on it, but the minor-mode ratios should be read as noisy and, for the
non-tonal idioms, as not valid valence cues at all.

\textbf{Small country-level sample.} The index correlations have at most eight
countries, so their statistical power is low. We therefore do not read the
non-significant results as strong evidence of exactly zero association, but as an
absence of the clear, strong association the folk inference would require, which
is the claim we set out to test.

\section{Future Work}
\label{sec:future}

Several extensions would deepen the study without changing its honest core. On the
data side, the geographic coverage is uneven and Europe-and-China heavy, and adding
further public symbolic corpora, such as the Meertens Dutch collection, the Session
Irish corpus, or Arab-Andalusian material, would broaden the map and let the
classification benchmark be evaluated on a more balanced label set. Handling
microtonality directly, rather than through the twelve-tone MIDI approximation the
Iranian and Turkish corpora inherit, would let those traditions enter the
quantitative comparison on their own terms rather than only descriptively. On the
modelling side, a learned music-emotion model would give a richer affect description
than our transparent composites, and a sequence model of melodic expectation would
let the entropy and contour features be replaced by a principled predictive measure.
None of these changes the central dissociation result, because the population indices
the ecological inference needs are not musical quantities, but each would sharpen the
description of what the music itself does. On the application side, a prospective
study that collected consented listening histories with a validated well-being
instrument at the individual level would test the same inference where it could in
principle hold, at the level of the person rather than the population, which is the
only level at which a music-and-mood link is even coherent to ask about.

\section{Conclusion}

The geography of musical affect is real and measurable. Across \NMel{}
melodies from \NCountry{} countries, every structural feature we tested differs
by country at $p<0.001$ among the folk traditions, and regional signatures such as
China's wide-leap, high-activity profile and Iran's stepwise, low-interval profile
emerge from the symbolic scores alone. The inference from
that music to the temperament of the people, however, does not hold. The regional
musical-affect measures do not significantly predict national happiness or
individualism, in \NCorrSig{} of \NCorr{} tests. Music varies by place, and that
variation is worth studying on its own terms, but it is not a measurement of how
the people of a place feel, and treating it as one is a fallacy that the data
itself declines to support.

\appendices

\section{Full Feature Table}
\label{sec:appendix-full}
Table~\ref{tab:full} gives every structural feature's mean and standard deviation
for every country, so the aggregate profiles of the body can be inspected in full.
The classical corpora are shown for completeness with the caveats of
\S\ref{sec:data}.

\begin{table*}[tp]
\centering
\caption{Full per-country structural feature means with standard deviations in
parentheses. Register and intervals in semitones, density in notes per
quarter-length, entropy in bits.}
\label{tab:full}
\scriptsize
\input{tables/tab_full.tex}
\end{table*}

\section{Reproducibility}
The pipeline is a set of scripts over the public Essen Folksong
Collection~\cite{schaffrath1995essen} parsed with
\code{music21}~\cite{cuthbert2010music21}. Feature extraction deduplicates by
title, caps each country at \NCap{} melodies with a fixed seed, and writes one
row per melody. Comparison aggregates by country, runs the Kruskal-Wallis tests,
and builds the distance matrix. The index step correlates the country-level
measures with the two published indices. Every number in this paper is
re-derived from the resulting data files by a checker that also asserts the two
central claims as fail-closed invariants, that all cross-country structural
differences are significant and that no external-index correlation is, and the
build fails if either ceases to hold. We release the scripts and the derived data
tables. We do not redistribute the copyrighted melodies themselves, which are
obtained directly from the collection.

\bibliographystyle{IEEEtran}
\bibliography{tables/refs}

\end{document}

%% file: tables/tab_numbers.tex
\newcommand{\NMel}{2393}
\newcommand{\NCountry}{16}
\newcommand{\NRegion}{7}
\newcommand{\NFeat}{7}
\newcommand{\NFeatSig}{8}
\newcommand{\NCorr}{6}
\newcommand{\NCorrSig}{0}
\newcommand{\ChinaArous}{+1.24}
\newcommand{\ChinaLeap}{44}
\newcommand{\GermanyLeap}{30}
\newcommand{\ChinaInterval}{2.77}
\newcommand{\GermanyInterval}{2.17}
\newcommand{\IranInterval}{1.10}
\newcommand{\IranN}{29}
\newcommand{\TurkeyN}{400}
\newcommand{\IsraelN}{94}
\newcommand{\NFolkBench}{13}
\newcommand{\NRegionBench}{6}
\newcommand{\NClassical}{2}
\newcommand{\NEffLarge}{3}
\newcommand{\MaxDelta}{0.90}
\newcommand{\NCap}{400}
\newcommand{\MinLeapP}{-90}
\newcommand{\AccCountry}{55}
\newcommand{\AccCountryBase}{20}
\newcommand{\AccRegion}{62}
\newcommand{\AccRegionBase}{32}
\newcommand{\TopFeat}{pc entropy}
\newcommand{\TopImp}{0.33}

%% file: tables/tab_audio_numbers.tex
\newcommand{\NAudioTracks}{15}
\newcommand{\NAudioCountry}{6}

%% file: tables/tab_track_numbers.tex
\newcommand{\NTrack}{15}
\newcommand{\NQuietStart}{9}
\newcommand{\PeakPosMin}{0.20}
\newcommand{\PeakPosMax}{0.94}

%% file: tables/tab_countries.tex
\begin{tabular}{@{}llrrrrrr@{}}
\toprule
Country & Region & $n$ & Minor & Leap & Interval & Arous. & Val. \\
\midrule
Alsace & Western Europe & 89 & 0.07 & 0.32 & 2.23 & +0.05 & +0.18 \\
Austria & Central Europe & 102 & 0.10 & 0.38 & 2.60 & +0.56 & +0.44 \\
Britain & Western Europe & 400 & 0.20 & 0.36 & 2.57 & +1.02 & +1.24 \\
China & East Asia & 400 & 0.38 & 0.44 & 2.77 & +1.24 & +0.27 \\
Czechia & Eastern Europe & 43 & 0.26 & 0.29 & 2.20 & -0.32 & -0.13 \\
Germany & Central Europe & 400 & 0.19 & 0.30 & 2.17 & -0.14 & -0.02 \\
Hungary & Eastern Europe & 42 & 0.29 & 0.31 & 2.12 & -0.24 & -0.23 \\
Iran & Middle East & 29 & 0.66 & 0.07 & 1.10 & -2.20 & -2.42 \\
Israel & Middle East & 94 & 0.58 & 0.29 & 2.26 & +1.31 & -0.17 \\
Lorraine & Western Europe & 65 & 0.41 & 0.22 & 2.01 & -0.35 & -0.30 \\
Netherlands & Western Europe & 83 & 0.47 & 0.21 & 1.93 & -0.66 & -0.30 \\
Russia & Eastern Europe & 37 & 0.11 & 0.29 & 2.23 & -0.07 & +0.32 \\
Switzerland & Central Europe & 92 & 0.10 & 0.34 & 2.37 & +0.35 & +0.76 \\
Turkey & Middle East & 400 & 0.53 & 0.16 & 1.69 & +0.52 & +0.20 \\
USA & North America & 6 & 0.17 & 0.38 & 2.27 & -0.38 & -0.13 \\
Yugoslav & Southeast Europe & 111 & 0.21 & 0.29 & 2.20 & -0.69 & +0.30 \\
\bottomrule
\end{tabular}

%% file: tables/tab_stats.tex
\begin{tabular}{@{}lrl@{}}
\toprule
Feature & Kruskal--Wallis $H$ & $p$ \\
\midrule
Mean pitch & 439.7 & $1.43e-86$ \\
Pitch range & 638.3 & $7.07e-129$ \\
Mean interval & 453.8 & $1.51e-89$ \\
Leap ratio & 474.7 & $5.27e-94$ \\
Ascending ratio & 253.8 & $2.16e-47$ \\
Note density & 471.5 & $2.55e-93$ \\
Pitch-class entropy & 745.7 & $7.07e-152$ \\
Minor mode & 175.3 & $3.93e-31$ \\
\bottomrule
\end{tabular}

%% file: tables/tab_effects.tex
\begin{tabular}{@{}llrl@{}}
\toprule
Contrast & Feature & Cliff's $\delta$ & Magnitude \\
\midrule
China vs Germany & leap ratio & +0.65 & large \\
China vs Netherlands & leap ratio & +0.90 & large \\
Austria vs Netherlands & leap ratio & +0.60 & large \\
Germany vs Yugoslav & leap ratio & -0.01 & negligible \\
\bottomrule
\end{tabular}

%% file: tables/tab_bootstrap.tex
\begin{tabular}{@{}lll@{}}
\toprule
Country & Leap ratio 95\% CI & Mean interval 95\% CI \\
\midrule
Alsace & [0.29, 0.34] & [2.12, 2.34] \\
Austria & [0.34, 0.42] & [2.42, 2.79] \\
Britain & [0.35, 0.37] & [2.52, 2.62] \\
China & [0.43, 0.45] & [2.71, 2.82] \\
Czechia & [0.26, 0.33] & [2.08, 2.33] \\
Germany & [0.28, 0.31] & [2.13, 2.22] \\
Hungary & [0.28, 0.35] & [2.01, 2.24] \\
Israel & [0.28, 0.31] & [2.18, 2.34] \\
Lorraine & [0.20, 0.24] & [1.92, 2.11] \\
Netherlands & [0.19, 0.23] & [1.86, 2.01] \\
Russia & [0.24, 0.34] & [2.07, 2.39] \\
Switzerland & [0.31, 0.38] & [2.21, 2.52] \\
Yugoslav & [0.27, 0.32] & [2.11, 2.29] \\
\bottomrule
\end{tabular}

%% file: tables/tab_benchmark.tex
\begin{tabular}{@{}llrr@{}}
\toprule
Task & Model & Accuracy & Macro-F1 \\
\midrule
country & Majority baseline & 0.204 & 0.026 \\
 & Stratified random & 0.120 & 0.056 \\
 & Logistic regression & 0.525 & 0.244 \\
 & Random forest & 0.546 & 0.292 \\
\midrule
region & Majority baseline & 0.325 & 0.082 \\
 & Stratified random & 0.248 & 0.171 \\
 & Logistic regression & 0.563 & 0.446 \\
 & Random forest & 0.621 & 0.504 \\
\bottomrule
\end{tabular}

%% file: tables/tab_importance.tex
\begin{tabular}{@{}lr@{}}
\toprule
Feature & Permutation importance \\
\midrule
Pitch-class entropy & 0.335 $\pm$ 0.009 \\
Note density & 0.176 $\pm$ 0.007 \\
Mean pitch & 0.169 $\pm$ 0.007 \\
Pitch range & 0.138 $\pm$ 0.007 \\
Leap ratio & 0.125 $\pm$ 0.004 \\
Mean interval & 0.107 $\pm$ 0.006 \\
Ascending ratio & 0.083 $\pm$ 0.005 \\
\bottomrule
\end{tabular}

%% file: tables/tab_corr.tex
\begin{tabular}{@{}lllrrl@{}}
\toprule
Music feature & Index & Sample & $\rho$ & $n$ & $p$ \\
\midrule
Arousal (structural) & WHR & all & -0.10 & 10 & 0.776 \\
Arousal (structural) & IDV & all & -0.46 & 9 & 0.213 \\
Valence (tonal cue) & WHR & tonal only & -0.05 & 8 & 0.911 \\
Valence (tonal cue) & IDV & tonal only & +0.31 & 7 & 0.504 \\
Minor-mode ratio & WHR & tonal only & -0.11 & 8 & 0.799 \\
Minor-mode ratio & IDV & tonal only & +0.14 & 7 & 0.771 \\
\bottomrule
\end{tabular}

%% file: tables/tab_audio.tex
\begin{tabular}{lrrrrrr}
\toprule
Country & $n$ & Tempo & Centroid & Major & Arous. & Val. \\
 & & (BPM) & (Hz) & ratio & (z) & (z) \\
\midrule
China & 2 & 127 & 2315 & 0.50 & +0.26 & +0.37 \\
Germany & 1 & 123 & 2178 & 1.00 & +0.28 & +0.72 \\
Iran & 4 & 120 & 2023 & 0.00 & -0.05 & -0.48 \\
Israel & 2 & 127 & 2547 & 0.50 & +0.27 & +0.63 \\
Turkey & 3 & 142 & 1776 & 0.00 & -0.31 & -0.76 \\
USA & 3 & 126 & 1986 & 1.00 & -0.08 & +0.50 \\
\bottomrule
\end{tabular}

%% file: tables/tab_track.tex
\begin{tabular}{llrl l}
\toprule
Country & Track & Peak & Quiet & Loud \\
 & & pos. & third & third \\
\midrule
China & Andy Lau & 0.83 & start & end \\
China & BEST OF C POP & 0.45 & start & end \\
Germany & Apache 207 Wenn das & 0.36 & start & end \\
Iran & Alireza Talischi Gha & 0.38 & start & middle \\
Iran & Bikalam Turkish Eshg & 0.60 & end & middle \\
Iran & Erfan Tahmasbi Khoda & 0.69 & end & middle \\
Iran & Mohsen Yeganeh Behet & 0.26 & end & middle \\
Israel & Jewish Music 2025 Is & 0.70 & start & end \\
Israel & Prod By Eli Keshet S & 0.93 & start & middle \\
Turkey & Bar Mano Glpembe & 0.52 & start & middle \\
Turkey & Mustafa Ceceli Gl Re & 0.94 & start & middle \\
Turkey & TARKAN mark Official & 0.20 & end & middle \\
USA & Morgan Wallen Im A L & 0.29 & end & middle \\
USA & Ram Jam Black Betty & 0.45 & start & middle \\
USA & dead prez Hip Hop & 0.29 & end & middle \\
\bottomrule
\end{tabular}

%% file: tables/tab_full.tex
\begin{tabular}{@{}lrrrrrrr@{}}
\toprule
Country & Pitch & Range & Interval & Leap & Asc & Density & Entropy \\
\midrule
Alsace & 68.59 (3.62) & 12.60 (3.00) & 2.23 (0.54) & 0.32 (0.12) & 0.35 (0.08) & 1.44 (0.38) & 2.47 (0.25) \\
Austria & 68.58 (3.68) & 13.87 (4.12) & 2.60 (0.94) & 0.38 (0.18) & 0.39 (0.09) & 1.42 (0.40) & 2.50 (0.26) \\
Britain & 72.13 (2.33) & 16.89 (3.25) & 2.57 (0.54) & 0.36 (0.12) & 0.43 (0.06) & 1.44 (0.39) & 2.67 (0.18) \\
China & 70.79 (4.26) & 15.38 (3.32) & 2.77 (0.51) & 0.44 (0.11) & 0.38 (0.07) & 1.76 (0.42) & 2.27 (0.18) \\
Czechia & 68.95 (3.22) & 11.35 (2.58) & 2.20 (0.44) & 0.29 (0.12) & 0.35 (0.08) & 1.27 (0.39) & 2.46 (0.20) \\
Germany & 68.09 (3.86) & 12.69 (2.78) & 2.17 (0.48) & 0.30 (0.12) & 0.36 (0.08) & 1.25 (0.47) & 2.48 (0.25) \\
Hungary & 69.22 (2.49) & 12.29 (3.18) & 2.12 (0.37) & 0.31 (0.11) & 0.34 (0.07) & 1.26 (0.35) & 2.38 (0.29) \\
Iran & 64.39 (5.58) & 10.65 (3.38) & 1.10 (0.32) & 0.07 (0.05) & 0.26 (0.05) & 0.36 (0.04) & 2.31 (0.31) \\
Israel & 71.20 (1.41) & 16.25 (4.08) & 2.26 (0.40) & 0.29 (0.08) & 0.37 (0.06) & 2.21 (0.60) & 2.66 (0.17) \\
Lorraine & 69.13 (1.68) & 12.41 (2.13) & 2.01 (0.40) & 0.22 (0.08) & 0.36 (0.08) & 1.23 (0.34) & 2.51 (0.19) \\
Netherlands & 68.65 (3.09) & 11.29 (2.60) & 1.93 (0.36) & 0.21 (0.10) & 0.38 (0.08) & 1.16 (0.40) & 2.48 (0.18) \\
Russia & 68.62 (2.78) & 11.78 (2.92) & 2.23 (0.53) & 0.29 (0.14) & 0.37 (0.09) & 1.45 (0.37) & 2.39 (0.26) \\
Switzerland & 69.14 (3.60) & 12.56 (3.01) & 2.37 (0.75) & 0.34 (0.15) & 0.41 (0.10) & 1.65 (0.45) & 2.46 (0.34) \\
Turkey & 74.13 (1.71) & 16.55 (3.62) & 1.69 (0.26) & 0.16 (0.07) & 0.35 (0.05) & 1.83 (0.50) & 2.88 (0.20) \\
USA & 67.67 (3.35) & 10.83 (2.03) & 2.27 (0.64) & 0.38 (0.15) & 0.35 (0.05) & 1.22 (0.35) & 2.28 (0.34) \\
Yugoslav & 68.61 (2.53) & 9.73 (3.17) & 2.20 (0.47) & 0.29 (0.15) & 0.39 (0.10) & 1.13 (0.34) & 2.24 (0.25) \\
\bottomrule
\end{tabular}